\documentclass[superscriptaddress,showpacs,aps,twocolumn,prb,floatfix]{revtex4}
\usepackage{bm,amsmath,amssymb}
\usepackage{graphicx,color}

\begin{document}

\title{Kondo physics of magnetic adatoms on metallic surfaces 
when the onset of the surface conduction density of 
states crosses the Fermi level}

\author{J. Fern\'andez}
\affiliation{Centro At\'omico Bariloche, CNEA, 8400 S. C. de
Bariloche, Argentina}

\author{P. Roura-Bas}
\affiliation{Centro At\'omico Bariloche, CNEA, 8400 S. C. de
Bariloche, Argentina}

\begin{abstract}
We study the role of the onset of Shockley states, $D_s$, belonging to (111) surfaces of Cu, Ag and Au 
in the Kondo effect when a magnetic impurity is deposited on them. When $D_s$ approaches to the Fermi level, $E_F$, 
thing that can be done by compressing (stretching) the metallic sample, we found that most of the thermodynamic 
and dynamic properties of the impurity are affected in a non trivial way. We model the system by a generic Anderson impurity 
model and solve it by using the numerical renormalization group, NRG, technique.
In particular, the impurity contribution to magnetic susceptibility 
and entropy as a function of temperature exhibit negative values and goes to zero slowly 
in a logarithmic shape. 
Furthermore, we found a suppression of the spectral density weight at the Fermi level when $D_s\sim E_F$ even
in the Kondo regime. As a consequence, the conductance through the impurity is strongly reduced by near $25\%$
of the unitary value $2e^2/h$. Finally, we analyze these features in realistic systems like Co on Ag(111) 
reported in the literature.
\end{abstract}

\maketitle

\section{Introduction}\label{intro}

In solid state physics, Shockley states arise when solving the Schr\"{o}dinger equation in the 
context of nearly free electron models as a consequence of crystal termination. 
They are a common feature of (111) surfaces of noble metals.\cite{Gauyacq} 
In case of Ag, Cu and Au metals, the d-surface bands are well below the 
Fermi energy $E_F$, while the sp-surface ones are located around $E_F$.\cite{ibach} 

The nearly constant surface density of states (sDOS) abruptly starts below $E_F$
at  $D_s-E_F\approx -450$ meV for Cu,\cite{knorr} $D_s-E_F\approx -475$ meV for Au,\cite{cerce} and 
$D_s-E_F\approx -67$ meV for Ag.\cite{limot} 
From scanning tunneling microscope (STM) measurements, the corresponding steps have been observed.\cite{cerce,limot,liu}
Interestingly, the occupation of these surface states depends appreciably on particular variables such 
as temperature and the presence of adsorbed species, and more importantly by stretching the sample.
Decreasing temperature, the onset of Ag(111) sDOS moves monotonically towards $E_F$. \cite{morgenstern}
Stretching the Ag layers, induced by the film growth, shifts $D_s$ up in energy, even beyond the Fermi level. \cite{neuhold} 
Furthermore, the onset of sDOS can be changed by alloying the different noble metals at the surface. \cite{cerce,abd}.
In addition, the construction of a piezoelectric-based apparatus for applying continuously
tunable compressive and tensile strains to test samples is reported in Ref. \onlinecite{Mackenzie}. 
It can be used within a wide temperature range, including cryogenic ones.\cite{Mackenzie,Mackenzie-2}.
In particular, such a device can be used for moving $D_s$ continuously across $E_F$.

\begin{figure}[tbp]
\begin{center}
\includegraphics[clip,width=\columnwidth]{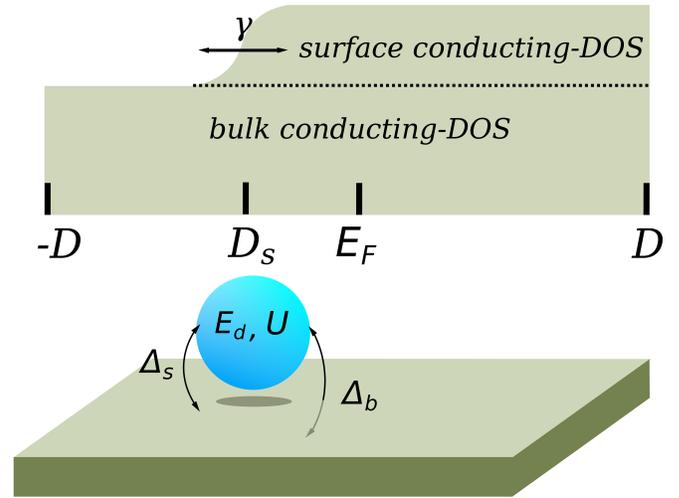}
\end{center}
\caption{Sketch of the setup. A magnetic impurity modeled as a single quantum dot with an energy level $E_d$ 
and Coulomb repulsion $U$ is deposited on the surface of a metal. The bulk density of the states 
extends from $-D$ to $D$ while the surface contribution abruptly starts at $D_S$. Both systems are coupled via 
the hybridization parameters $\Delta_b$ and $\Delta_s$. $E_F$ indicates the Fermi energy. }
\label{fig1}
\end{figure}

On the other hand, the above mentioned metal surfaces are often used to host magnetic impurities 
(see figure \ref{fig1}). 
Several experiments were made to study atomic impurities (such as Co or Mn) over such surfaces.
\cite{knorr,limot,li,madha,man,serrate-1,serrate-2, reso,li-2} Moreover, complex magnetic molecules were also deposited
on Ag, Cu and Au surfaces like FePc \cite{hiraoka,minamitani} among others.   
With help of low-temperature STM measurements, in these systems the Kondo effect have been studied.
In such measurements, the Kondo phenomena emerges trough a narrow Fano-Kondo 
antiresonance in the differential conductance $G(V)=dI/dV$, where $I$ is the current and $V$ 
the applied voltage.
 
The width of $G(V)$ is usually related with the Kondo temperature, $T_K$, and below it the magnetic
moment of the impurity is screened by the conduction electrons.\cite{Ujsaghy} 

The Kondo effect has the quality of universality in the sense that most of impurity properties display an universal 
shape once they are scaled by $T_K$. This fact makes $T_K$ 
the only relevant energy scale of the problem and its precise estimation is always desirable. 

Having mentioned that the onset of (111) surfaces can be move towards the Fermi energy and taking into account
that the Kondo effect is a low energy phenomena, the question of how it is affected when  $|D_s-E_F|\sim T_K$
raises. 
Interestingly, when $|D_s-E_F|\gg T_K$ the Kondo temperature as a function of $D_s$ was found to display 
a power law $T_K \simeq C |D_s-E_F|^{\alpha}$.\cite{joaquin-step}

In this paper we analyze the Kondo physics of a magnetic impurity on metallic surfaces 
when the onset of the surface conduction density of states crosses the Fermi level. 
We found that $T_K$ looses the power law dependence and more importantly, both thermodynamic and transport 
properties of the impurity in the Kondo regime are strongly modified in a non trivial way.

Modeling the system with an Anderson impurity Hamiltonian, we employ 
the numerical renormalization group (NRG) technique as implemented in the 
NRG LJUBLJANA open source code \cite{nrg-code}. 

In particular, when $|D_s-E_F| \le T_K$ we found unusual temperature dependencies in the impurity 
properties like negative values of entropy, $S_{imp}(T)$, and magnetic susceptibility, $\chi_{imp}(T)$.
Furthermore, since $D_s$ can be moved towards $E_F$, we also analyze the conductance, $G(T)$, 
through the impurity encountering a strong suppression of the conductance al low 
temperatures when the onset approaches to $E_F$. 
Finally and making contact with experiments, we study the conductance within the regime in which measurements
by Limot \textit{et al.} in Ref. \cite{limot} and by Moro-Lagares \textit{et al.} in Ref. \cite{serrate-2} 
were made.

Our results are important in order to predict deviations of the Kondo properties from its usual behaviors in systems 
where the onset of sDOS moves towards the Fermi energy. 

The paper is organized as follows. In Section \ref{model} we described the theoretical model under study. 
Numerical results and its analysis are given in Section \ref{results}. Finally, a summary and conclusions are 
presented in Section \ref{summary}.

\section{Model}\label{model}

The theoretical model we analyze in this paper, corresponding to the setup in figure \ref{fig1}, 
is defined by the following Hamiltonian

\begin{eqnarray}
H &=&\sum_{k\sigma }\varepsilon _{k}^{s}s_{k\sigma }^{\dagger }s_{k\sigma
}+\sum_{k\sigma }\varepsilon _{k}^{b}b_{k\sigma }^{\dagger }b_{k\sigma
}+\notag \\
&&E_{d}\sum_{\sigma }n_{d\sigma }+ 
U\sum d_{\uparrow }^{\dagger }d_{\uparrow }d_{\downarrow }^{\dagger
}d_{\downarrow }+\notag \\
&&\sum_{k\sigma }V_{k}^{s}[d_{\sigma }^{\dagger }s_{k\sigma }+%
\text{H.c.}]+\sum_{k\sigma }V_{k}^{b}[d_{\sigma }^{\dagger }b_{k\sigma }+\text{H.c.}].
\label{ham}
\end{eqnarray}

where the first two terms correspond to surface and bulk non-interacting conduction electrons
respectively. The operator $d_{\sigma }^{\dagger }$ creates an electron with spin $\sigma $ at 
the magnetic impurity with $n_{d\sigma}=d_{\sigma }^{\dagger}d_{\sigma}$. $E_{d}$ and $U$ are 
the energy level and Coulomb repulsion respectively.
The last two terms describe the hybridization between the impurity and conduction electrons.

As usual, we describe the bulk contribution to conduction electrons with a constant 
density of states, $\rho_b$, extended in a band-width from $-D$ to $D$. 
On the other hand, we model the surface contribution to the density of states including lifetime effects 
coming from quasiparticles interactions by a rounded-step function (see figure \ref{fig1}) 

\begin{equation}\label{rho-s}
\rho_s(\omega)=\frac{\rho_s}{\pi}\Big[ \frac{\pi}{2} + 
\mbox{arctan}\big( (\omega - D_s)/(\gamma/2) \big) \Big]\theta (D-\vert\omega\vert).
\end{equation}

that starts at $D_s$ and is extended to $D$ where $\gamma$ 
represents the inverse lifetime of the surface states. \cite{limot,lifetime,eigenenergy-zitko} 
Regarding the upper limit of sDOS, it was found to be of the order of $1$ eV in Au(111) \cite{ibach} and $0.2$ 
eV in Ag(111) \cite{liu}. In any case, we chose for simplicity the upper limit to coincide with the one 
of the bulk DOS, $D$. This choice is justified because it does not modified the low energy properties, 
near the Fermi level, which is the central point in our work.
To provide an order of magnitude of $\rho_b$ and $\rho_s$, they were found to be $0.135/$eV and $0.0466/$eV in 
case of Ag(111) respectively. \cite{reso}

The effect of the conduction bands on the impurity can be put in terms of an energy dependent 
hybridization within the range $-D$ to $D$

\begin{eqnarray}\label{delta}
\Delta(\omega)&&\equiv\pi\sum_{\nu k\sigma}\vert V_{k}^{\nu} 
                                       \vert^{2}\delta(\omega-\varepsilon _{k}^{\nu})\nonumber\\
              &&=\Delta_{s}\tilde{\rho}_s(\omega) +\Delta_{b}.
\end{eqnarray}

where $\Delta_{\nu}=\pi {\vert V^{\nu}\vert}^{2} \rho_{\nu}$ and $\tilde{\rho}_{\nu}=\rho_{\nu}(\omega)/\rho_{\nu}$
is the bare density of states of the band electrons, normalized to its constant value.

The retarded Green function of the impurity is given by
\begin{equation}
G_{d\sigma }(z)=\frac{1}{z-E_{d}-\Sigma _{\Delta\sigma }(z)-\Sigma _{U\sigma }(z)}%
,  \label{gd}
\end{equation}%
where $z=\omega + i0$ and $\Sigma _{U\sigma }(z)$ and $\Sigma _{\Delta\sigma}(z)$ are the interacting 
(due to the Coulomb repulsion $U$) and the non-interacting (due to the one body hybridization $\Delta(\omega)$)
contributions to the self-energy respectively.

The non-interacting part of the self-energy is defined by the hybridization function

\begin{equation}\label{sigma0}
\Sigma_{\Delta\sigma}(\omega)=\Gamma(\omega)-i\Delta(\omega),
\end{equation}

where the real part $\Gamma(\omega)$ can be obtained from $\Delta(\omega)$ by means of a Kramers-Kronig transformation.

As we mentioned in the Introduction, we solve the Hamiltonian in Eq. (\ref{ham}) by using the NRG technique. 
In addition to the usual z-averaging \cite{z-averaging} and sigma-trick \cite{sigma-trick} 
refinements when calculating thermodynamics and dynamical properties, the NRG-ljubljana uses an 
improved discretization scheme \cite{discretization} that allows NRG method to handle with energy dependent
hybridizations \cite{eigenenergy-zitko}. As usual, the half bandwidth of the bulk conduction electrons
is taken as the unit of energy, $D=1$ and the Fermi energy is chosen to be $E_F=0$. Furthermore, 
we have used a renormalization parameter $\Lambda=2$ and a number of different values of $z$, $N_z=16$,
in all NRG calculations.

\section{Results}\label{results}
\subsection{Sharp-step surface DOS}

We start this section considering that the surface density of states can be modeled by a step function, that is without
the broadening introduced by the inverse lifetime, $\gamma=0$. 
This case was considered before for a fixed value of $\vert D_s\vert\gg T_K$. 
\cite{serrate-2,reso,joaquin-step,eigenenergy-zitko,remir} 
As we will show in this work, in the regime $\vert D_s\vert\ll T_K$ the step shape of sDOS contains the key ingredients related with
the role of the onset in the Kondo physics.

\subsubsection{Kondo temperature as a function of $D_s$}

It will be instructive to present a brief summary of the main results of Ref. \onlinecite{joaquin-step} 
regarding the variation of $T_K$ with $D_s$ for $\vert D_s\vert\gg T_K$.
When the onset of the surface contribution to conduction density of states  
is near the Fermi energy, the Kondo temperature as a function of $D_s$ was found to display 
a power law $T_K \simeq C |D_s|^{\alpha}$, where the exponent $\alpha$ depends on the 
relative intensities between surface and bulk hybridizations
with the impurity, $\Delta_s/\Delta_b$, and the ratio between on-site and
Coulomb energies $E_d/U$ as well as the sign of $D_s$ and $C$ encloses the other dependences.  
This power law was obtained from a 
poor man's scaling (PMS) \cite{hewson,and} approach to the effective Kondo model and confirmed numerically
by using the Non-Crossing Approximation (NCA) \cite{hewson,bickers}. 
The validity of the above dependence applies for small values of $|D_s| \ll D$ but they are limited 
to $|D_s| \gg T_K$.  Unfortunately, neither PMS nor NCA can described the regime 
$|D_s| \le T_K$. In case of PMS, the renormalization group procedure ceases to be valid 
\cite{hewson} while in case of NCA, there are inaccurate results when dealing with small energies as compared 
with the Kondo one \cite{bickers}.

Specifically, the Kondo scale was found to depend
on the onset $D_s$ in the following way
\begin{eqnarray}
T_{K} &\simeq &A|D_s|^{\eta }D^{1-\eta }\exp \left[ \frac{\pi
E_{d}(E_{d}+U)}{2U(\Delta _{b}+\Delta _{s})}\right] ,  \notag \\
\eta  &=&\frac{\Delta _{s}}{(\Delta _{b}+\Delta _{s})}\left(1+\frac{E_{d}}{U}\right),%
\text{ for }D_s<0, \label{scaling-neg}
\end{eqnarray}%
and
\begin{eqnarray}
T_{K} &\simeq &B(D_s)^{\zeta }D^{1-\zeta }\exp \left[ \frac{\pi
E_{d}(E_{d}+U)}{2U\Delta _{b}}\right] ,  \notag \\
\zeta  &=&\frac{\Delta _{s}E_{d}}{\Delta _{b}U},\text{ for }D_s>0.
\label{scaling-pos}
\end{eqnarray}%

Both equations are valid in the limit of $\vert D_s \vert \gg T_K$ and when
charge fluctuations are limited only to two configurations, $0\le n_d \le 1$ for $E_d<\infty$, $U\rightarrow\infty$ 
and $1\le n_d \le 2$ for $E_d+U<\infty$, $\{E_d,-U\}\rightarrow-\infty$. 
The exponents $\eta$ and $\zeta$ were confirmed by NCA calculations for several values of the ratio $\Delta_s/\Delta_b$.

Within our NRG calculations, we get the Kondo temperature from the thermodynamic properties of the Hamiltonian
according to Wilson's definition, $ k_B T_K \chi_{imp}(k_B T_K) / (g\mu_B)^2 = 0.07$, being 
$\chi_{imp}(k_B T)$ the impurity contribution to the magnetic susceptibility as a function of temperature 
\cite{review-wilson}. Through the rest of the manuscript we set $g\mu_B=1$ and $k_B=1$.

\begin{figure}[tbp]
\begin{center}
\includegraphics[clip,width=\columnwidth]{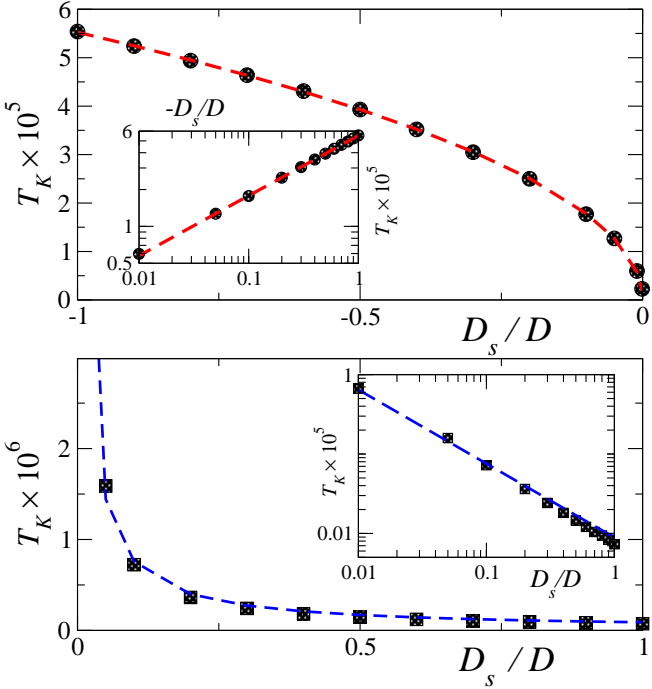}
\end{center}
\caption{(Color online) Kondo temperature as a function of $D_s$ for $E_d=-0.04D$, $U\rightarrow \infty$ (top panel) and 
$E_d+U=0.04D$, $\{E_d,-U\}\rightarrow-\infty$ (bottom panel). In both cases $\Delta_b=\Delta_s=0.5D$.
Similar values of $E_d$ and $E_d+U$ were used in Ref. \onlinecite{joaquin-step}.
Fitting parameters: $A=5.52\times10^{-5}$, $\eta=0.49$, $cc=0.9999$, 
$B=8.9\times10^{-8}$, $\zeta=-0.94$, $cc=0.9998$. The insets show the results in a logarithmic scale.}
\label{fig2}
\end{figure}

In figure \ref{fig2} we show the results for $T_K$ as a function of $D_s$ in the selected case of 
$\Delta_s=\Delta_b$. 
The top panel displays the case of having $U\rightarrow\infty$. From Eq. (\ref{scaling-neg}), a value of 
$\eta=1/2$ is expected. Fitting the NRG results with a power law of the form $A\vert D_s \vert^\eta$ 
for negative values of $D_s$ we 
obtain an exponent $\eta=0.49$. On the other hand, the lower panel shows the results in case of 
$E_d + U < \infty$ while $\{E_d,-U\}\rightarrow-\infty$ for which an exponent $\zeta=-1$ for positive values 
of $D_s$ is expected. 
We obtain $\zeta=-0.94$ from the fitting. In both cases the accuracy of the fittings is confirmed 
by a correlation coefficients (cc) near to unity with an error of $10^{-4}$. 
The present results, also limited to $T_K\ll \vert D_s \vert$, verifying the corresponding ones presented in Ref. 
\onlinecite{joaquin-step}. Furthermore, other ratios of $\Delta_s/\Delta_b$ (not shown) were studied and also 
agree with the PMS and NCA results.

\begin{figure}[tbp]
\begin{center}
\includegraphics[clip,width=\columnwidth]{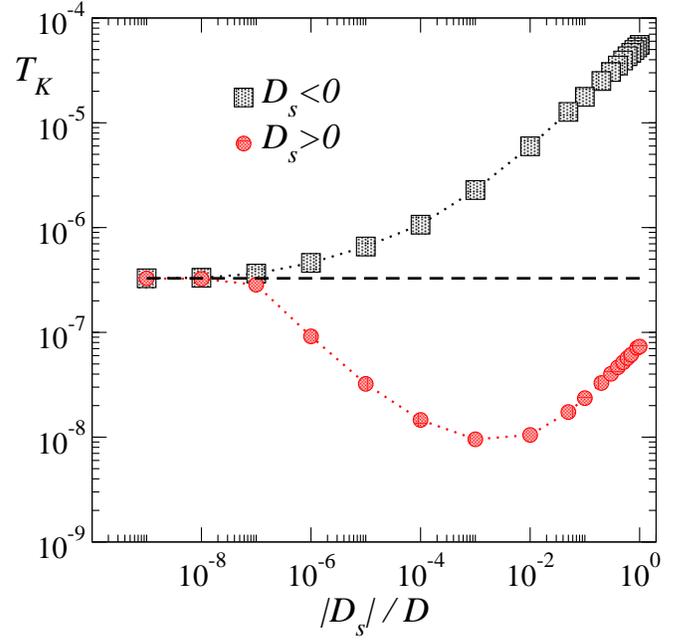}
\end{center}
\caption{(Color online) Kondo temperature as a function of $D_s$ for the same set of model 
parameters that in top panel of 
figure \ref{fig2}. 
Dashed line stands for $D_s = 0$ representing an energy scale $T^{\ast}=3.3 \times 10^{-7}D$.}
\label{fig3}
\end{figure}

As we mentioned, the range $T_K \ge \vert D_s \vert$ is not accessible neither by PMS nor NCA.
In figure \ref{fig3} we show the values of $T_K$ when $D_s$ approaches to the Fermi energy from negative values 
in case of $U\rightarrow\infty$ and finite on-site energy $E_d$. 

We can see from the plot that the scaling law in Eq. (\ref{scaling-neg}) breaks down as soon as $-D_s$ 
approaches to $T_K$. 
Furthermore, $T_K$ seems to be constant in the range $\vert D_s \vert \ll T_K$ and 
saturates at a value $T^\ast$ when $D_s=0$ (dashed line). 
The corresponding values for positive $D_s$, although they do not obey any scaling, are present
in order to show that a continuity function $T_K(D_s)$ is obtained when $D_s$ crosses the Fermi energy. 
Finally, we found (not shown) a similar deviation of the power law in case of Eq. (\ref{scaling-pos}).

\subsubsection{Thermodynamic properties in the range $\vert D_s \vert \ll T_K$}

In what follows we turn our attention to the analysis of the 
Kondo physics through the impurity contribution to thermodynamic properties in the range 
$\vert D_s \vert \ll T_K$ and in particular for $D_s=0$.

\begin{figure}[tbp]
\begin{center}
 \includegraphics[clip,width=\columnwidth]{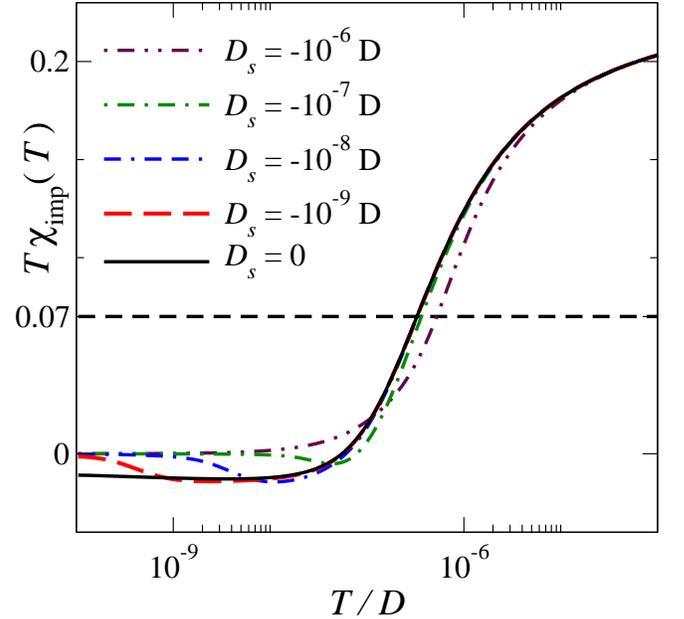}
\end{center}
\caption{(Color online) Impurity magnetic susceptibility as a function of temperature for the same set of model parameters 
that in figure \ref{fig3} and for different values of $D_s/D$. Black dashed constant line indicates the 
criteria for defining the Kondo temperature.}
\label{fig4}
\end{figure}

Interestingly, when $\vert D_s \vert < T_K$, the impurity magnetic susceptibility times temperature, 
$T\chi_{imp}(T)$, approaches to zero from negative values and very slowly. 
This unusual low-temperature behavior is shown in figure \ref{fig4} for 
the same set of model parameters that figure \ref{fig3}.
Notice that for finite values of $D_s$, the magnitude of $T\chi_{imp}(T)$ vanishes at sufficient 
low temperature. In particular, for $D_s=0$, $T\chi_{imp}(T)$ approaches to zero
in the limit of $T\rightarrow0$. 
From the results of figure \ref{fig3}, we do not expect a significant variation of the impurity properties for 
$D_s > 0$.

\begin{figure}[tbp]
\begin{center}
 \includegraphics[clip,width=\columnwidth]{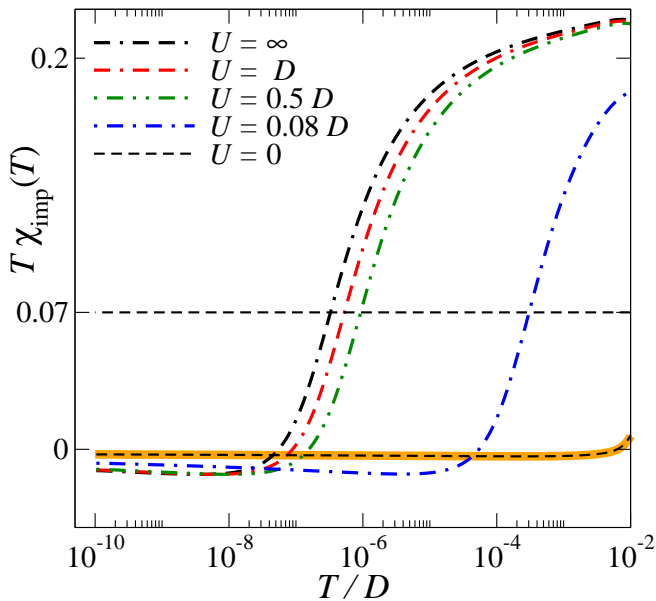}
 \end{center}
\caption{(Color online) Impurity magnetic susceptibility as a function of temperature in case of $D_s=0$ and for 
different values of the 
Coulomb repulsion $U$ and for $E_d=-0.04D$. Solid orange line corresponds to the analytic 
result for $U=0$ according to Eq. (\ref{non-interacting}).}
\label{fig5}
\end{figure}

\begin{figure}[tbp]
\begin{center}
 \includegraphics[clip,width=\columnwidth]{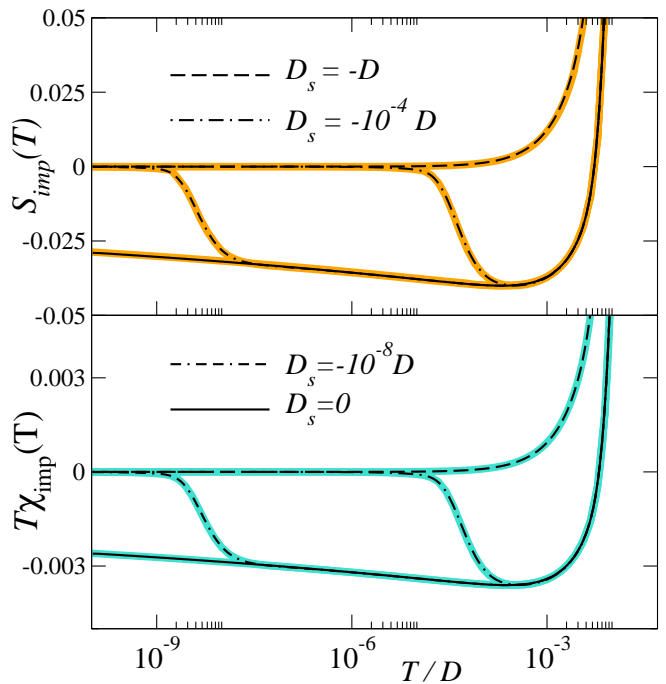}
 \end{center}
\caption{(Color online) In thick lines, $S_{\textmd{imp}}(T)$ (top panel) and $T\chi_{\textmd{imp}}(T)$ (bottom panel) 
results for $D_s=U=0$ from Eqs. \ref{non-interacting}. Black light lines correspond to NRG calculations.}
\label{fig6}
\end{figure}

When analyzing thermodynamic properties of the impurity, we also notice that similar low temperature
features also appears in the impurity contribution to the entropy, $S_{imp}(T)$. That is, $S_{imp}(T)$
goes to zero at low temperature from negative values (see top panel of figure \ref{fig6}). 

We remain the reader that the impurity contribution to $\chi(T)$ and $S(T)$ are given by 
$\chi_{imp}(T)=\chi_{tot}(T)-\chi_{tot}^{(0)}(T)$ and $S_{imp}(T)=S_{tot}(T)-S_{tot}^{(0)}(T)$
where $\chi_{tot}^{(0)}(T)$ ($S_{tot}^{(0)}(T)$) denotes the system without impurity.
\cite{review-wilson, review-bulla} 
Therefore, even 
when $\chi_{tot}(T)$ ($S_{tot}(T)$) and $\chi_{tot}^{(0)}(T)$ ($S_{tot}^{(0)}(T)$) are separately 
positive, the impurity contribution can be negative. 

One may wonder if the Coulomb repulsion takes any role in this behavior. In figure \ref{fig5} we fix the onset of 
the surface density of states in $D_s=0$ and $E_d=-0.04D$ and analyze $\chi_{imp}(T)$ for several values of
$U$ from the non-interacting case $U=0$ to $U\rightarrow\infty$.

In fact, they are already present for the non-interacting case $U=0$. The 
distinctive feature of the model Hamiltonian in Eq. (\ref{ham}) is enclosed in the one-body self-energy in 
Eq. (\ref{sigma0}), so in the following we analyze the magnetic susceptibility and entropy for the non-interacting
model.

The impurity magnetic susceptibility and entropy for the non-interacting version of the Hamiltonian in Eq. (\ref{ham})
are given by the following expressions

\begin{eqnarray}\label{non-interacting}
S_{imp}(T) &=& \frac{2}{\pi T}\mbox{Im}\int_{-\infty}^{+\infty} d\omega f'(\omega) \omega \mbox{ln}
                              G_{0d}^{-1}(\omega) \\
T\chi_{imp}(T) &=& \frac{1}{2\pi}\mbox{Im}\int_{-\infty}^{+\infty} d\omega f'(\omega) (1-2f(\omega)) \mbox{ln}
                              G_{0d}^{-1}(\omega),\nonumber
\end{eqnarray}

where $G_{0d}^{-1}(\omega)=\omega-E_{d}-\Sigma_{\Delta}(\omega)$ is the non-interacting  
electron impurity Green function independently of the spin.

Both magnitudes were calculated in the case of $D_s=0$ and its results as a function of temperature are 
displayed in figure \ref{fig6} represented by solid lines. 
Clearly, the same features that in figures \ref{fig4} and \ref{fig5} are present.
Although it is not necessary, we take the opportunity to benchmark the NRG when calculating 
the same thermodynamic impurity contribution for $U=0$. 
As it is clear from the figure, the results obtained are on-top of the analytical ones. This accuracy 
when dealing whit energy dependent hybridizations is due to the improved discretization scheme 
detailed in Ref. \cite{discretization}.

The low-temperature behavior of Eq. (\ref{non-interacting}) is given by the corresponding low-energy form of the 
one-body self-energy, which in the limit of $D_s=0$ is 

\begin{eqnarray}\label{low-energy-sigma0}
 \Sigma_{\Delta}(\omega)\sim \frac{\Delta _{s}}{\pi }\ln \Big\vert \frac{\omega }{D}\Big\vert 
-i\Delta(\omega).
\end{eqnarray}

Note the divergent form at low energies. 
Then, the impurity entropy and magnetic susceptibility becomes

\begin{eqnarray}
S_{imp}(T) &\sim& -\frac{(1/2)\mbox{ln}4}{\mbox{ln}(1/2T)} + O(T^2)\nonumber\\
T\chi_{imp}(T) &\sim& -\frac{(1/8)}{\mbox{ln}(1/2T)} + O(T^2).
\end{eqnarray}

Remarkably, the usual Fermi liquid properties for $\vert D_s\vert\le T_K$ are recovered at low enough temperature and 
for the special case in which the onset of the surface DOS coincides with the Fermi energy, they are reach 
logarithmically slowly.

Here we would like to mention the work of Mitchell and Fritz, Ref. \onlinecite{mitchell}, in which they
analyze the Kondo effect in graphene with vacancies. In spite of being very different 
systems, the one studied by Mitchell and co-workers and the present one, 
they also found negative impurity contributions to magnetic susceptibility and entropy at low temperatures
independently of the value of the Coulomb repulsion. 
In that case, the responsible of such behavior is attributed to a logarithmically divergent 
\textit{imaginary} part of the one-body hybridization $\Sigma_{\Delta}$ coming from a zero mode in graphene
with defects. In our case, we have a logarithmically divergent \textit{real} part of the one-body 
hybridization coming from a non-analyticity in the conduction DOS. In any case, both hybridization shapes 
give rise similar low-temperature features in the impurity properties.

\begin{figure}[tbp]
\begin{center}
 \includegraphics[clip,width=\columnwidth]{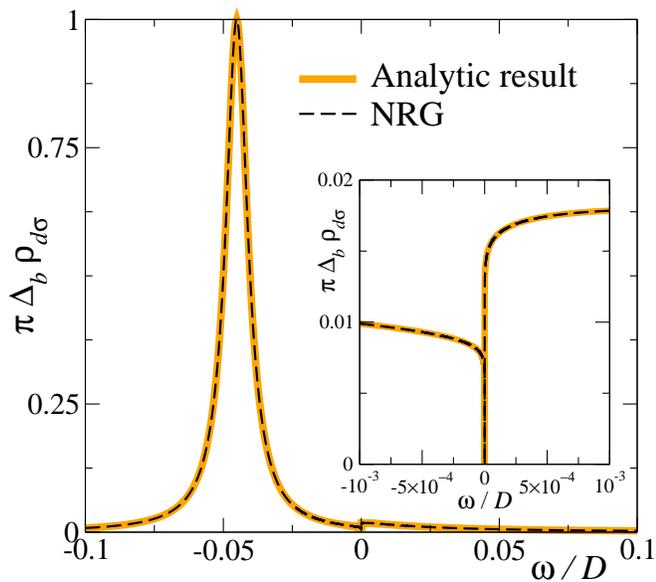}
 \end{center}
\caption{(Color online) Non-interacting impurity spectral density $\rho_{0d}(\omega)$ for $D_s=0$ 
obtained from Eq. (\ref{gd}) with $\Sigma _{U\sigma }(z)=0$ and NRG with $U=0$. 
Other parameters as in figure \ref{fig2}.}
\label{fig7}
\end{figure}

\subsubsection{Spectral density for $D_s = 0$}

Since $\Sigma_{\Delta}(\omega)$ becomes infinite for $D_s=\omega=0$, it is expected that the 
impurity spectral density, $\rho_{0d}(\omega)=- \mbox{Im} G_{0d}(\omega)/\pi$, vanishes at the 
Fermi energy. This is explicitly shown in figure \ref{fig7} in which both, the analytical and 
NRG calculations are again on-top one each other.

\begin{figure}[tbp]
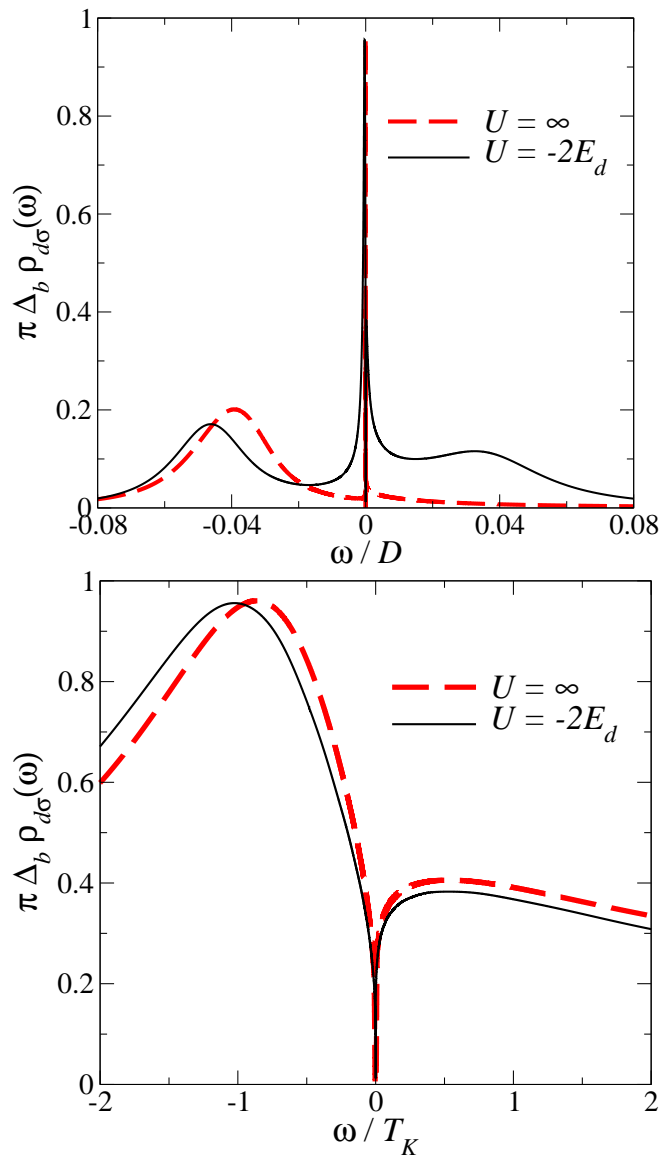

\begin{center}
\includegraphics[clip,width=\columnwidth]{fig8a.eps}
\includegraphics[clip,width=\columnwidth]{fig8b.eps}
\end{center}
\caption{(Color online) Top panel: spectral densities $\rho_{d\sigma}$ as a function of frequency for $D_s=0$ 
and $U=-2Ed$ (solid line) and $U=\infty$ (dashed line). $E_d=-0.04$ and $\Delta_s=\Delta_b=0.005$.
Lower panel: low energy region in units of $\omega/T_K$ being $T_K(U=-2Ed)=3.04\times10^{-4}D$ and 
$T_K(U=\infty)=3.40\times10^{-7}D$. Calculations were done at $T=10^{-3} T_K$. }
\label{fig8}
\end{figure}

As we have shown, the Coulomb repulsion is not responsible for this effect so one may wonder in which way this 
logarithmic energy dependence affects the Kondo effect. 
Once the interaction $U$ is turned on and the temperature is lowered, the spectral weight at the 
Fermi level should be increased due to the appearance of the Kondo peak.\cite{hewson} 
Therefore, a competition between both effects
is expected.
In figure \ref{fig8} the spectral density is shown for two selected values of the Coulomb repulsion, 
$U=-2E_d$ and $U=\infty$ and for $D_s=0$. The NRG calculations were done at $T=10^{-3}T_K$ being $T_K$ the 
corresponding one for each value of $U$ accordingly to figure \ref{fig5}.
Several aspects can be discuss from these results. In a first place, the top panel shows the spectral 
density in its extended frequency region. The solid line has a value of the Coulomb interaction in such a way that
$2E_d+U=0$, which is the usual condition for particle-hole symmetry. This symmetry is clearly broken due
to the asymmetry in the hybridization function in Eq. (\ref{delta}) (see also the cartoon of the conduction DOS 
in figure \ref{fig1}). 
While the lower charge transfer peak, 
locate at energy $\omega\sim E_d$ has a width of the order of $4\Delta_b$ \cite{capa,width} the upper one, 
locate at $\omega\sim E_d+U$ has a width of the order of $4(\Delta_b+\Delta_s)$ and, as consequence, 
its intensity is reduced. 
In the $U=\infty$ case, a similar width of $4\Delta_b$ in the unique charge transfer peak is shared.
The central peak correspond to the Kondo resonance. In a similar manner that the width of the Fano-Kondo 
antiresonance is related with $T_K$, the width of the Kondo peak does. 
In relation with the low energies features of the spectral density, 
the bottom panel of figure \ref{fig8} shows a detail of the Kondo resonance. Once the energy dependence in
both curves is shown in units of $T_K$, they are quite similar as a consequence of the universality of 
the Kondo phenomena.

\begin{figure}[tbp]
\begin{center}
\includegraphics[clip,width=\columnwidth]{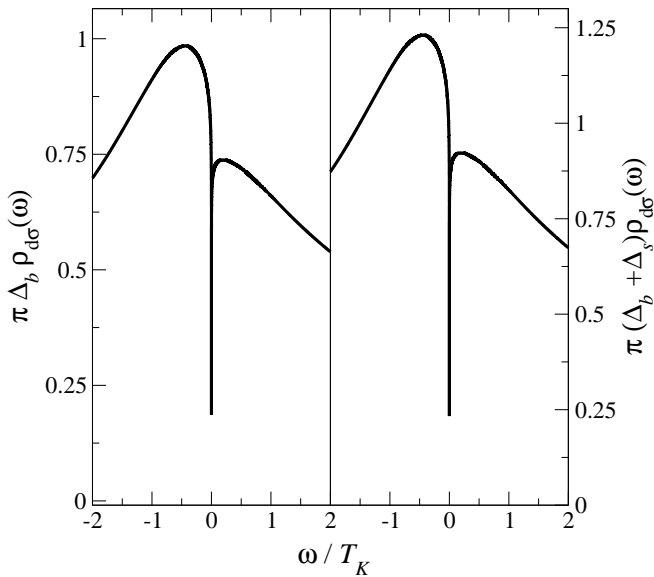}
\end{center}
\caption{(Color online) Spectral density $\rho_{d\sigma}$ in the 
low energy region in units of $\omega/T_K$ ($T_K=3\times10^{-5}D$) 
for $D_s=0$ and $\Delta_s=0.25\Delta_b=0.00125$. $U=-2Ed$ with $E_d=-0.04$.
Left panel: $\rho_{d\sigma}$ scaled by $\pi\Delta_b$. Right panel: 
$\rho_{d\sigma}$ scaled by $\pi( \Delta_b+\Delta_s )$ .}
\label{fig9}
\end{figure}

The logarithmically divergence of the one-body hybridization splits the Kondo resonance into two pieces 
around the Fermi level. Furthermore, a suppression of the spectral weight at $\omega=E_F$ is 
observed independently of the value of the Coulomb repulsion. Note that for negative energies closes to 
the Fermi one, the Kondo peak seems to agree with the usual Friedel sum rule $\rho_{d\sigma}(0)=1/\pi\Delta$,
with $\Delta=\Delta_b$. On the other hand, since the sDOS is already turned on for positive energies,
the hole contribution of the Kondo resonance, $\omega\gtrsim 0$, seems to be related with the total 
contribution $\Delta=\Delta_b+\Delta_s$.
We have confirmed the last statement by plotting the spectral density for another relation between 
$\Delta_b$ and $\Delta_s$. In particular we choose the ratio $\Delta_s/\Delta_b=1/4$ which has found to be 
the lower limit for the surface contribution to the total hybridization in Co impurities deposited on Ag(111)
surface.\cite{serrate-2} What we observed from the plot is that while the negative portion of the Kondo 
resonance approaches to one when scaled by $\pi\Delta_b$, the positive one reaches the same value but when
it is scaled by $\pi( \Delta_b+\Delta_s )$. Note that for the selected parameters the Kondo resonance is slightly
asymmetric with more weight in the negative region, so it is closer to one 
for negative energies than for the positive ones.

In any case, the usual Friedel sum rule does not apply since $\rho_{d\sigma}(0)\sim 0$. 
Instead, as we will see, the weight of the spectral density at the Fermi energy is consistent with the generalized 
Friedel sum rule. \cite{loig, lan}

\subsubsection{Conductance through the impurity and occupation}

The reduction of the spectral weight al low energies affects the transport properties of the model.

\begin{figure}[tbp]
\begin{center}
\includegraphics[clip,width=\columnwidth]{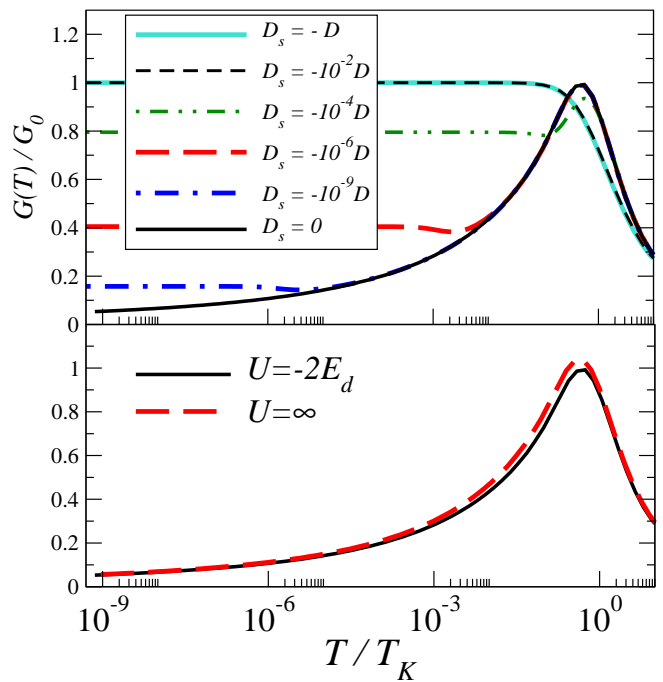}
\end{center}
\caption{(Color online) Equilibrium conductance in units of $G_0=2e^2/h$ as a function of 
temperature in units of $T_K$. Top panel: Several values of $D_s$ for the case $U+2E_d=0$ with
$U=0.08D$. Bottom panel: $D_s=0$ for the cases $U+2E_d=0$ with $U=0.08D$, black solid curve, 
and $U\rightarrow\infty$, red dashed line.}
\label{fig10}
\end{figure}

The equilibrium conductance $G(T)$ directly depends on $\rho_{d\sigma}(\omega)$ \cite{win}

\begin{eqnarray}\label{conductance}
G(T) = G_0\frac{\pi}{2}\sum_{\sigma}\int_{-\infty}^{+\infty}d\omega ~ 
 [-f'(\omega)] \Delta(\omega)\rho_{d\sigma}(\omega),
\end{eqnarray}

where $G_0=2e^2/h$ is the quantum of conductance.
In particular, a strong reduction of the equilibrium conductance, $G(T)$, 
at low temperatures is expected.
In figure \ref{fig10} we show the NRG calculations for the conductance as a function of temperature.
In the top panel we have fixed $U+2E_d=0$ with $U=0.08D$ while varying the values of the onset $D_s$.
When $D_s=-D$ the surface contribution to the hybridization agrees with the bulk one and the conductance displays
the usual behavior reaching the maximum value for this problem of $G_0$ for temperatures $T\ll T_K$.
For smaller values of $|D_s|$ but still large as compared with the corresponding $T_K$, the 
shape of the conductance is not affected. However, when $-D_s$ is of the order of $T_K$ 
($\sim 10^{-4}D$), $G(T)$ 
saturates at a reduced fraction of $G_0$. 
The reduction of the saturated conductance grows as $D_s$ approaches to zero. 
In the limiting case of $D_s=0$, the conductance vanishes at low temperatures in a logarithmic shape.

In the bottom panel of figure \ref{fig10} we show the obtained values of $G(T)$ for $D_s=0$ and two
different values of the Coulomb repulsion, $U+2E_d=0$ with $U=0.08D$, black solid curve, 
and $U\rightarrow\infty$, red dashed line. As it is clear from the results, when scaling the temperatures
by the corresponding Kondo one, both curves display an identical temperature dependence as expected 
from the universality of the Kondo effect.

\begin{figure}[tbp]
\begin{center}
\includegraphics[clip,width=\columnwidth]{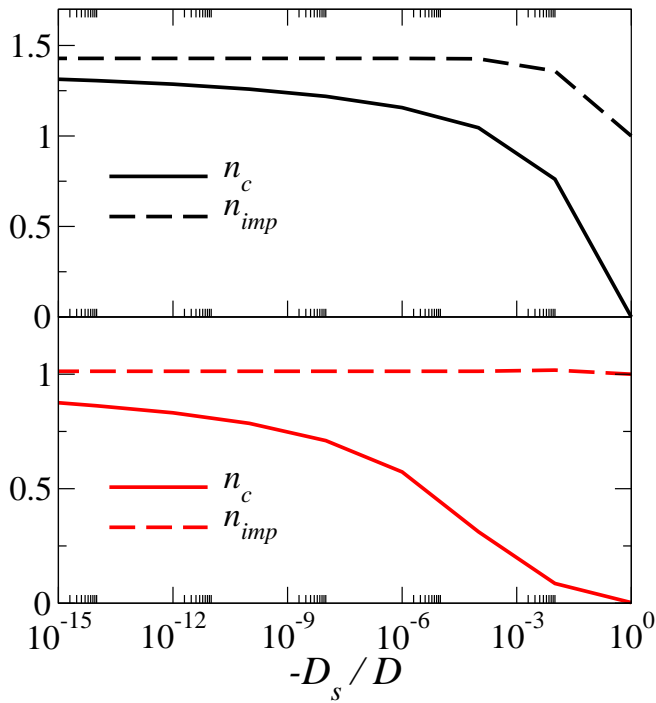}
\end{center}
\caption{(Color online) $n_c$ and $n_{imp}$ as a function of $D_s$ for $U+2E_d=0$. Top panel: $U=0$. 
Bottom panel: $U=0.08D$. In both cases we have used $\Delta_s=\Delta_b=0.005D$.}
\label{fig11}
\end{figure}

The results of the intermediate plateaus in the conductance can be understood in terms of the generalized 
Friedel sum rule.\cite{loig, lan} 

\begin{eqnarray}\label{generalized-friedel}
G(T)/G_0 &=& {\mbox{sin}}^2 \Big( \frac{\pi}{2} (n_{imp}-n_c) \Big)\\
         &=& \pi(\Delta_b+\Delta_s)\rho_{d\sigma}(0)\nonumber,
\end{eqnarray}

where $n_{imp}$ is the impurity occupation and 

\begin{eqnarray}\label{nc}
n_c = -\mbox{Im}\sum_{\sigma} \int_{-\infty}^{E_F} \frac{d\omega}{\pi} ~ 
  G_{d\sigma}(\omega)\frac{ \partial\Delta(\omega)}{\partial\omega},
\end{eqnarray}

is related with the change of the charge in the conduction band as a consequence of the 
presence of the impurity.
In the usual case of a flat hybridization, this term vanishes due to 
$\frac{ \partial\Delta(\omega)}{\partial\omega}\sim0$. However in our case we expect its
influence to be rather large when $D_{s}\sim E_{F}$. 
For the non-interacting model, we evaluate Eq. (\ref{nc}) and together with the results of 
the impurity occupation, Eq. (\ref{generalized-friedel}) can be verified. In the top panel
of figure \ref{fig11} we show the results of $n_c$ and $n_{imp}$ at zero temperature and 
as a function of $D_s$ is the case of $U+2E_d=0$ with $U=0$. While the impurity population is
rather constant, it is clear that $n_c$ approaches to $n_{imp}$ as $\vert D_s \vert $ is
decreased. Therefore, both $G(T\rightarrow 0)$ and $\rho_{d\sigma}(\omega\rightarrow 0)$
vanish when $D_s\rightarrow 0$. 
In the bottom panel of figure \ref{fig11} we show the NRG results for $n_c$ and $n_{imp}$ 
at temperature $T=10^{-3}T_K$ and as a function of $D_s$ is case of $U+2E_d=0$ with 
$U=0.08D$. As we have previously mentioned, particle-hole symmetry is slightly broken 
(see figure \ref{fig8}) and therefore the impurity occupation is very close, but not exactly, to one.
However, $n_c$ strongly changes from zero ($D_s=-D$) to one ($D_s=0$). In this case,
when interactions are turning on, we obtain $n_c$ not directly from Eq. (\ref{nc}) 
but from the more accurate values of the low temperature conductance according to 
Eq. (\ref{generalized-friedel}). Note that in the case of $D_s=0$ the result $n_c=n_{imp}$
is only reached asymptotically for $T\rightarrow 0$, see figure \ref{fig10}.

\subsection{rounded-step surface DOS}

We ending the present work by considering finite lifetime effects, $\gamma$, leading to a smoother 
onset of the band edge as shown in Eq. (\ref{rho-s}).

This is specially important to make contact with real systems. In particular with the experiment of 
Limot and co-workers \cite{limot} in which the authors perform scanning tunneling spectroscopy measurements
of magnetic (Co) and 
non-magnetic (Ag) atoms on Ag(111) and Cu(111) and analyze a bound state that appears after impurity deposition 
near the onset of sDOS. In case of Co/Ag(111) and Co/Cu(111) a dip corresponding to a Fano-Kondo antiresonance 
appears at the Fermi level. 
For a quantitative analysis of the experiment, the authors 
model the surface DOS by including lifetime effects. 

Therefore, it is interesting to see how rapidly the logarithmic shapes of the analyzed properties 
in this paper are blurring in the presence of $\gamma$. 
When a Co atom is deposited on a clean Ag(111) surface, the measure $T_K\sim 83 K$ is found and the 
data is adjusted with $\gamma=7meV=81.2K$ and therefore $\gamma\sim T_K$.
On the other hand, when a Co atom is deposited on Cu(111) the values were $T_K\sim 63 K$ and 
$\gamma=24meV=278.4K$ which leads to a ratio $\gamma/T_K\sim 4.4$.

\begin{figure}[tbp]
\begin{center}
 \includegraphics[clip,width=\columnwidth]{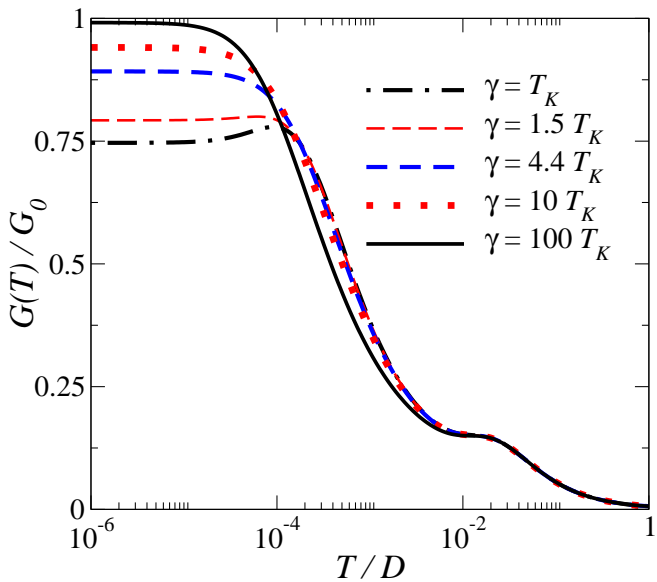}
\end{center}
\caption{Conductance as a function of temperature in case of $D_s = 0$ for several values of the inverse lifetime $\gamma$
in units of $T_K$. As a reference, we use $T_K$ corresponding to the black solid line in figure \ref{fig8}. 
$\gamma=T_K$ and $\gamma=4.4T_K$ represent Co/Ag(111) and Co/Cu(111) respectively in Limot's experiment, 
while $\gamma=1.5T_K$ represents Co/Ag(111) in Moro-Lagares' experiment.}
\label{fig12}
\end{figure}

In figure \ref{fig12} we show the NRG results for the conductance as a function of temperature when
the surface hybridization is given by Eq. (\ref{rho-s}) and for several values of the ratio $\gamma/T_K$.
Here, we do not expect to make a quantitative description of each system mentioned above, but provide 
an analysis of what will be the effect of $\gamma$ given the measure values of $T_K$. To this end, we have
chosen a generic set of model parameters, $E_d$, $U$, $\Delta_b$ and $\Delta_s$ and compared the 
effect of $\gamma$ against the obtained $T_K$. Provided that in the Kondo regime, the only relevant 
energy scale is in fact $T_K$ and all magnitudes display a universal dependence when scaled by it, 
as for instance the spectral density in figure \ref{fig8} (lower panel) or the conductance in figure
\ref{fig10}, the results in figure \ref{fig12} are quite general. 
The case of $\gamma/T_K=1$ represents Co/Ag(111) in Limot's experiment. Clearly, when the onset of the 
surface states coincides whit the Fermi energy, even in the case of having a lifetime in the DOS, the 
conductance is reduced significantly, near a factor $25\%$ of the usual ideal value of $G_0$. 
The case of $\gamma/T_K=4.4$ represents Co/Cu(111) with also an important reduction of the low temperature
conductance. For comparison, we also show the case $\gamma/T_K=10$ for which the effect of the step is still
present. Only when $\gamma/T_K\ge100$ the conducting DOS is smooth enough to get $n_c\sim0$ and the ideal
conductance can be reached. Note however that here we are not including other degrees of freedom in the system 
that contribute to enlarge the inverse lifetime in sDOS. 
Such contributions should be include in order to fully analyze the Kondo properties.

\begin{figure}[tbp]
\begin{center}
 \includegraphics[clip,width=\columnwidth]{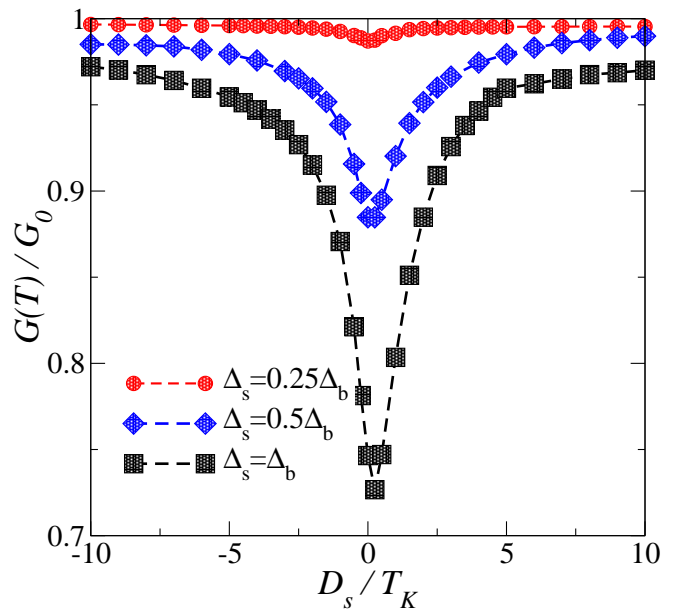}
\end{center}
\caption{ Saturated conductance extracted from the low temperature plateaus in figure \ref{fig12} 
as a function of $D_s$ for several values of the ratio $r_{sb}=\Delta_s/\Delta_b$. $\gamma=T_K$ as Co/Ag(111) in 
Limot's experiment where $T_K$ 
corresponding to the black solid line in figure \ref{fig8}.}
\label{fig13}
\end{figure}

In addition to the Limot's experiment, the Moro-Lagares and co-workers one in Ref. \onlinecite{serrate-2}, 
also study the Kondo phenomena that appears after deposition of a Co atom on a clean Ag(111) surface. 
In that work, the main result is a quantification of the role of the surface state in the Kondo effect and provides
a lower bound for the ratio of hybridizations $r_{sb}=\Delta_s/\Delta_b=1/4$. 

Regarding $T_K$, the Fano-Kondo line shape is also analyzed and they found $T_K=52K$ which increases the 
ratio $\gamma/T_K\sim 1.56$ as compared with Limot's work. The discrepancy 
between the Kondo temperature estimated by Limot \textit{et al.} and Moro-Lagares \textit{et al.} is explained in Ref. 
\onlinecite{diego} and will not be discuss here. 
For this ratio between $\gamma$ and $T_K$, the low temperature plateau in the conductance is reduced by near $20\%$.

In any case, the low temperature dependence of $G(T)$ is modified by the presence of the onset $D_s$ even
if a non vanishing inverse lifetime $\gamma$ is incorporated when describing the surface conducting electrons. 

Another way to observed the influence on transport measurements when moving the onset of sDOS, 
we plot in figure \ref{fig13} the saturated conductance as a function of $D_s$. 
When $\gamma=T_K$, as Co/Ag(111) in Limot's experiment, a dip centered in $D_s=0$ is clearly visible. 
The intensity of the dip depends on the ratio $r_{sb}$. Increasing $r_{sb}$ results
in an increment of the dip intensity. 
Once $\vert D_s \vert \gg T_K$, the conductance approach towards $G_0$ independently of $r_{sb}$.
Regarding the values of the ratio $r_{sb}$, R. \v{Z}itko in Ref. \onlinecite{eigenenergy-zitko} 
arguments in favor of $r_{sb}=1$ while the experiment of Moro-Lagares, as we have mentioned, results in $r_{sb}\ge1/4$. 
Note that from the results shown if figure \ref{fig10}, for $\gamma=0$ the dip reaches its maximum intensity,
$G(D_s=0)\rightarrow0$ independently of $r_{sb}$.
Therefore, for the whole realistic values of $r_{sb}$, even if $\gamma$ is present, evidence in 
the conductance through the impurity when $D_s$ crosses the Fermi level should be present.

\section{Summary}\label{summary}
We have analyzed thermodynamic and dynamical properties of a magnetic ad-atom deposited on a metal surface 
which contains a rounded step in the surface contribution to the total density of states of conduction electrons.
The model represents realistic setups for magnetic ad-atoms,  like Co, on (111) surface of 
Cu, Ag and Au. In such metals, a two-dimensional surface band ( Shockley states ) starts at an energy $D_s$ 
moderately below the Fermi level.
As we mentioned in the Introduction, $D_s$ can be  move continuously.

Within this framework, we have studied the low temperature properties inside the Kondo regime 
as a function of $D_s$, in particular when it crosses $E_F$.

Firstly, we confirm by means of exact numerical renormalization group calculations, the power law
Kondo temperature dependence on $D_s$ found recently by using Poor man's scaling arguments and Non Crossing
Approximation calculations in Ref. \onlinecite{joaquin-step}. 
After that, we focus on the physics that emerges from $D_s=0$. We investigate several magnitude of interest,
like impurity contribution to magnetic susceptibility and entropy, spectral density and equilibrium conductance.
In all magnitudes, the step in surface DOS plays an important role when $D_s$ approaches to $E_F$.
Interestingly, the magnetic susceptibility and entropy as a function of temperature 
exhibit negative values and goes to zero slowly in a logarithmic shape. This results becomes independently of
the impurity parameters such as the Coulomb repulsion being already present in the non-interacting case. 

Secondly, we examine the impurity spectral density as well as the conductance. Both are larger affected by $D_s$.
The results are interpreted in terms of the generalized Friedel sum rule in which 
the change of the charge in the conduction band, $n_c$, strongly depends on $D_s$.

Finally, we make emphasis in the relevance of our study not only for academic reasons but also in real
experiments. We analyze the situation in the experiment made by Limot \textit{et al.} in Ref. \onlinecite{limot}
and also in the one perform by Moro-Lagares and co-workers one in Ref. \onlinecite{serrate-2}. 

In both experiments a dip in the $dI/dV$ representing the Fano-antiresonance is found at bias voltage $V=0$. 
The dip is a clear manifestation of the Kondo resonance in the spectral density of Co. 
Although the results of figure \ref{fig8} represent the limit case of $D_s\sim 0$, the shape and intensity of such dips 
are already affected by the step in the surface DOS even for finite values of $D_s$. In particular, 
we have found that, if $D_s$ is moved towards $E_F$, measurable effects emerge. When the low-temperature 
conductance is plotted as a function of $D_s$, thing that can now be experimentally done \cite{Mackenzie}, 
we show that a dip should be present around  $D_s\sim0$.  Therefore, we expect that our work stimulates 
further research on the area of magnetic ad-atom on metallic surfaces.

\section*{Acknowledgments}
We thank A. A. Aligia and L. O. Manuel and I. Hamad for useful comments and careful reading of 
the manuscript. This work was sponsored by PIP 364 of CONICET (Argentina) and PICT xxx.
2013-1045 of the ANPCyT.

\end{document}